\documentclass[aps,prd,peprint,superscriptaddress,nofootinbib,12pt]{revtex4-2}

\usepackage[utf8]{inputenc}       
\usepackage[T1]{fontenc}          
\usepackage{amsmath,amssymb,amsfonts}  
\usepackage{mathtools}            
\usepackage{physics}              
\usepackage{bm}                   
\usepackage{tensor}               
\usepackage{slashed}              
\usepackage{braket}               
\usepackage{cancel}               
\usepackage{graphicx}             
\usepackage[dvipsnames]{xcolor}   
\usepackage[colorlinks=true, linkcolor=blue, citecolor=blue, urlcolor=blue]{hyperref} 
\usepackage{comment}              
\usepackage{enumitem}             
\usepackage{booktabs}             
\usepackage{float}                
\usepackage{caption}              
\usepackage{subcaption}           
\usepackage{natbib}

\begin{document}

\title{First Principles Quantization of a Non-Conservative Scalar Field}

\author{Kauship Saha}
\email{kauships24@iitk.ac.in}
\affiliation{Department of Physics, Indian Institute of Technology Kanpur, Kanpur 208016, India}

\author{Sandeep Aashish}
\email{aashish@iitp.ac.in}
\affiliation{Department of Physics, Indian Institute of Technology Patna, Patna, Bihar 801106, India}

\begin{abstract}
We present the first-principles quantization of a damped scalar field within the framework of classical action principle of non-conservative systems using doubled dynamical variables. We consider a non-conservative potential term constructed to describe a linear damping of the scalar field for quantization using canonical and path-integral formalisms, and derive the two-point Green's function along with the spectral function, which are consistent with known results from the well-known in-in formalism.
\end{abstract}

\date{\today}

\maketitle

\section{Introduction}
The principle of stationary action (or Hamilton’s principle) \cite{Goldstein2002} underpins virtually all of physics, providing a unified framework for deriving equations of motion—from simple classical systems (e.g., the harmonic oscillator) to quantum field theories with infinitely many degrees of freedom. However, it applies only to conservative (time-symmetric) systems; in non‐conservative (time-asymmetric) settings — even for a linear dissipative oscillator — the Hamilton principle fails \cite{Bauer1931}. Early on, Bateman \cite{Bateman1931} addressed linear dissipation by introducing an extra degree of freedom evolving backward in time, so that the combined system conserved energy and remained compatible with Hamilton’s principle. However, this approach has received limited attention due to its restriction to linear dissipation. Related doubling‐of‐variables schemes have modeled radiation reaction in charged and massive bodies \cite{Staruszkiewicz1970,Jaranowski1997}, but these approaches prioritize reproducing known damping forces over deriving them from first principles.

The difficulties intensify in the quantum domain. Caldirola and Kanai first quantized a single dissipative oscillator via a time‐dependent Hamiltonian, but this led to non‐unitary evolution, time‐dependent commutators, and apparent violations of the uncertainty principle \cite{Caldirola1941,Kanai1948}. Subsequent modifications restored the uncertainty relation \cite{Um:2002ab}, yet most canonical or path‐integral formulations still either break unitarity or violate uncertainty \cite{Dekker1981}. A multitude of quantization schemes for discrete dissipative systems has since emerged \cite{Banerjee:2001yc,Blacker:2021nto,Deguchi:2020ljd,Gitman:2006cp,Riseborough:1985zz,Dito:2005xt}, but they generally lack a foundation in first principles. The Lindblad formulation~\cite{Manzano:2020yyw, Nathan2020} provides a powerful mathematical framework for modeling open quantum systems, and numerous studies~\cite{Bertlmann:2006fn, McDonald:2023mbk, Park:2024lsa, Guff:2023xzf} have employed it to describe dissipation and particle decay; however, it is fundamentally an effective approach, rather than a strictly first-principles. Even in field theory, non-conservative interactions still pose significant challenges. Gaset et al.~\cite{Gaset2021} introduced a k-contact Lagrangian formalism that embeds dissipation into the action. While promising for classical field theories, it is currently limited to regular first-order Lagrangians and lacks a clear quantization scheme. Moreover, Boyanovsky et al. \cite{Boyanovsky1995} demonstrated that conventional perturbation theory fails when dissipation is present. Effective field theory(EFT) ~\cite{Crossley:2015evo,Boyanovsky:2015xoa,Fosco:2007nz,McDonald2023}, approaches to dissipation often introduce non-conservative terms phenomenologically. While successful in specific cases, these methods typically lack derivation from first principles. These issues underscore the need for a true first-principles description of non-conservative dynamics, without ad hoc tweaks to the Lagrangian or Hamiltonian.

A significant step toward a first-principles formulation was taken by Galley~\cite{Galley2013}, whose method is explicitly inspired by the in-in or Schwinger-Keldysh formalism~\cite{Schwinger1961, Keldysh:1964ud,Sieberer2016} and the Feynman–Vernon influence functional~\cite{FEYNMAN1963118}. The central idea is to double the degrees of freedom so that one set of degrees of freedom evolves forward in time while the other set evolves backward in time, and upon imposing a so-called "physical limit" only the forward evolving variables contribute to the equation of motion. This framework enables the construction of an action whose variation yields the correct equations of motion for general non-conservative systems. Several studies have adopted this approach to explore diverse non-conservative phenomena, including dissipative radiation reaction~\cite{Kalin:2022hph, Aashish:2016osv}, extensions of Noether’s theorem to non-conservative systems~\cite{Martinez-Perez:2017tng, Martinez-Perez:2015aza}, and quantization schemes for discrete non-conservative systems~\cite{Mahapatra:2023rzz, Mahapatra:2025sbu}. More recently, further extension of this approach to resolve certain subtleties and develop a full Hamiltonian theory has been carried out by Aykroyd et. al. \cite{aykroyd2025}. Although the parallels with quantum field theoretic Schwinger-Keldysh formalism is well acknowledged in the literature \cite{Sieberer2016, Baggioli2020}, validating the consistency of the Galley's classical framework through a first-principles quantization of non-conservative field theories remains unexplored. In this paper, we therefore undertake the quantization of a simple model of damped scalar field and reproduce known results from Schwinger-Keldysh formalism.  

We begin by briefly reviewing classical field theory in Galley's formulation \cite{Galley2013} in Sec.~\ref{sec: CLASSICAL THEORY OF NON-CONSERVATIVE FIELDS}, and introduce the non-conservative scalar field model that reproduces the damped Klein–Gordon equation. In Sec.~\ref{sec: THE TWO-POINT GREEN'S FUNCTION FROM FEYNMAN PATH INTEGRAL} we derive the two-point retarded and advanced Green's functions via the path-integral formulation and discuss the spectral function of the dissipative scalar field. Finally, in Sec.~\ref{sec: CANONICAL QUANTIZATION} we perform the canonical quantization of the dissipative scalar field and analyze the excitations that appear in the presence of dissipation.

\section{Classical theory of non-conservative fields}
\label{sec: CLASSICAL THEORY OF NON-CONSERVATIVE FIELDS}
In the framework of Ref.~\cite{Galley2013}, the number of degrees of freedom is doubled to transform the action principle from a boundary value problem to an initial value problem, without changing the fundamental structure of the action formalism. At the classical level, for a general field theory with non-conservative dynamics, the field $\Phi$ is replaced by a pair of new fields i.e. $\Phi \rightarrow (\Phi_1, \Phi_2)$, where $\Phi_1$ and $\Phi_2$ are `virtual' fields that evolve forward and backward in time respectively. The action, which is a functional of $\Phi_1$ and $\Phi_2$, is defined as
\begin{equation}
    S[\Phi_a] = \int d^4x\,\Lambda\bigl(\Phi_a,\partial_\mu\Phi_a\bigr)\,,
    \quad a = 1,2\,,
    \label{eg:action}
\end{equation}
where the Lagrangian density $\Lambda$ is
\begin{equation}
    \Lambda\bigl(\Phi_a,\partial_\mu\Phi_a\bigr)
    = \mathcal{L}\bigl(\Phi_1,\partial_\mu\Phi_1\bigr)
    - \mathcal{L}\bigl(\Phi_2,\partial_\mu\Phi_2\bigr)
    + \mathcal{K}\bigl(\Phi_{1,2},\partial_\mu\Phi_{1,2}\bigr)\,
    \label{eg:lagrangian}
\end{equation}
By design, the dynamics of action (\ref{eg:action}) reduces to the known (``correct") equation of motion in the ``physical limit" where the distinction between the forward and backward evolving fields vanishes while preserving the non-conservative nature captured by the coupling or interaction between the two. In Eq.~\eqref{eg:lagrangian}, the first two terms constitute the conservative part of the Lagrangian density, while the final term, $\mathcal{K}(\Phi_{1,2},\partial_{\mu}\Phi_{1,2})$, represents the non‑conservative interaction.

It is standard practice to work with a more convenient set of fields $(\Phi_+,\Phi_-)$ defined as a linear combination of $(\Phi_1,\Phi_2)$:
\begin{equation}
  \Phi_+ = \frac{1}{2}\bigl(\Phi_1 + \Phi_2\bigr)\,, 
  \qquad
  \Phi_- = \Phi_1 - \Phi_2.
  \label{eg:newphitrans}
\end{equation}
At the classical level, the physical limit is then defined as $\Phi_+\to\Phi$ and $\Phi_-\to 0$.

A similar transformation also applies to the canonical momenta, given by
\begin{equation}
    \Pi_+ = \frac{1}{2}\bigl(\Pi_1 + \Pi_2\bigr)\,, 
    \qquad
    \Pi_- = \Pi_1 - \Pi_2\,.
    \label{eg:new-pitran.}
\end{equation}
From Eq.~\eqref{eg:new-pitran.}, the conjugate momenta are obtained as:
\begin{equation}
    \Pi_\pm = \frac{\partial \Lambda}{\partial(\partial_0 \Phi_\mp)}\,.  
    \label{eg:pm-momenta}
\end{equation}

Under the variation $\Phi_a(x)\longrightarrow \Phi_a(x,\epsilon)=\Phi_a(x,0)+\epsilon\,\eta_a(x),$ where $\epsilon<<1$ and $\eta_a$ are arbitrary variational functions obeying the boundary conditions of Galley's variational principle \cite{Galley2013,Galley:2014wla}. Expanding the action in Eq~\eqref{eg:action} to first order in \(\epsilon\) yields,
\begin{equation}
\begin{aligned}
S[\Phi_a(\epsilon)]
&= \int d^4x\;\Bigg\{\Lambda\big|_{\epsilon=0}
+ \epsilon\,\eta_a(x)\left.\frac{\partial\Lambda}{\partial\Phi_a}\right|_{\epsilon=0} \\
&\qquad\qquad\qquad
+ \epsilon\,\partial_\nu\eta_a(x)\left.\frac{\partial\Lambda}{\partial(\partial_\nu\Phi_a)}\right|_{\epsilon=0}
+ \mathcal{O}(\epsilon^2)\Bigg\},
\quad  a=+,-\,
\end{aligned}
\label{eq:action_expansion}
\end{equation}
Using Galley's boundary conditions together with Hamilton's principle \cite{Galley2013,Galley:2014wla}, the action is made stationary $\frac{\partial S}{\partial\epsilon}\big|_{\epsilon=0}=0$ which yields,
\begin{equation}
    \partial_\mu \frac{\partial\Lambda}{\partial(\partial_\mu \Phi_a)} = \frac{\partial\Lambda}{\partial\Phi_a}\,, \quad a = +, -\,.  
    \label{eg:eq_of_motion_before_Pl} 
\end{equation}

By applying the physical limit (PL) to Eq.~\eqref{eg:eq_of_motion_before_Pl}, only the variables labeled with "$+$" survive. As a result, we obtain the generalized Euler–Lagrange equation:

\begin{equation}
\label{eg:generalized_EL}
\partial_\mu\frac{\partial\mathcal{L}}{\partial(\partial_\mu\Phi)}
- \frac{\partial\mathcal{L}}{\partial\Phi}
=
\left[
    \frac{\partial\mathcal{K}}{\partial\Phi_-}
    - \partial_\mu\!\left(\frac{\partial\mathcal{K}}{\partial(\partial_\mu\Phi_-)}\right)
\right]_{\mathrm{PL}}.
\end{equation}

The transition from configuration space to phase space is achieved through a Legendre transformation, which leads to the Hamiltonian formulation:

\begin{equation}
    \mathcal{H}(\Phi_a, \Pi_a) = \Pi_+ \Phi_- + \Pi_- \Phi_+ - \mathcal{L}(\Phi_a, \partial_\mu \Phi_a) 
    \label{eg:hamiltonian density}
\end{equation}

This expression represents the Hamiltonian density in this formalism. The total Hamiltonian is obtained by integrating Eq.~\eqref{eg:hamiltonian density} over three-dimensional space. A more detailed discussion of the Hamiltonian formalism is available in Ref. \cite{aykroyd2025}.

\subsection{The model}
\label{sec: DISSIPATION IN SCALAR FIELD THEORY}
In the study of damped real scalar theory, the equation of motion is modified by a linear damping term, as discussed in Ref.~\cite{CartasFuentevilla2023, Carrillo-Ibarra:2013roa, Alves-Junior:2023glp}.
\begin{equation}
    \partial_t^2\phi(x,t)-\nabla^2 \phi(x,t) + m^2 \phi(x,t) + \gamma \partial_t\phi(x,t) = 0
    \label{eg:Kg dissipation}
\end{equation}
In the framework of Ref.~\cite{Galley2013}, the Lagrangian density from which Eq.~\eqref{eg:Kg dissipation} is derived requires doubling the degrees of freedom and introducing a non-conservative interaction term, $\mathcal{K}(\phi_{1,2},\partial_\mu\phi_{1,2})$. We therefore construct a Lagrangian $\Omega(\phi_{1,2},\partial_{\mu}\phi_{1,2})$ given by,
\begin{equation}
\begin{aligned}
\Omega(\phi_{1,2},\partial_\mu\phi_{1,2})
=\;& \frac{1}{2}\,\partial_\mu\phi_1\,\partial^\mu\phi_1
     - \frac{1}{2}\,m^2\phi_1^2 - \biggl(\frac{1}{2}\,\partial_\mu\phi_2\,\partial^\mu\phi_2 - \frac{1}{2}\,m^2\phi_2^2\biggr) + \mathcal{K}(\phi_{1,2},\partial_\mu\phi_{1,2}),
\end{aligned}
\label{eg:dissipativelagrangian}
\end{equation}
where
\[
\mathcal{K}(\phi_{1,2},\partial_\mu\phi_{1,2})
= -\frac{\gamma}{2}\,(\phi_1 - \phi_2)\,(\partial_0\phi_1 + \partial_0\phi_2)\,.
\label{}
\]
In Eq.~\eqref{eg:dissipativelagrangian}, the inclusion of the non-conservative interaction makes the Lagrangian density non-covariant. This behavior is expected in this framework, as dissipation breaks time-reversal symmetry, in accordance with the second law of thermodynamics and the arrow of time. For further reading on dissipation in field theory and the arrow of time, see Ref.~\cite{Vitiello:2001ut, Jafari:2017sbh, Jafari:2018rfk, Bodeker:2022ihg, Refaei:2015vfa, Ipek:2014saf, Berera:1998gx}.

The Lagrangian density can be expressed in terms of the $\phi_\pm$ variables of Eq.~\eqref{eg:new-pitran.} as
\begin{equation}
\Omega(\phi_\pm,\partial_\mu\phi_\pm)
= \partial_\mu \phi_-\,\partial^\mu \phi_+
  - m^2 \,\phi_+ \,\phi_-
  - \gamma\,\phi_-\,\dot{\phi}_+\,
  \label{eg:pm L}
\end{equation}
The conjugate momenta corresponding to the $\phi_\pm$ variables, as defined in Eq.~\eqref{eg:pm-momenta}, are given by
\begin{equation}
\begin{aligned}
    \Pi_+ &= \frac{\partial \Omega}{\partial(\partial_0 \phi_-)} = \partial_0 \phi_+\,,\\
    \Pi_- &= \frac{\partial \Omega}{\partial(\partial_0 \phi_+)} = \partial_0 \phi_- - \gamma\,\phi_-\,
    \label{eg:+-momentum of scalar field}
\end{aligned}
\end{equation}
The corresponding Hamiltonian of this model is given by, 
\begin{equation}
    H = \int d^3 x \left( \Pi_+ \Pi_- + \nabla \phi_+ \cdot \nabla \phi_- + m^2 \phi_+ \phi_- + \gamma \Pi_+ \phi_- \right)
    \label{eq:H_density}
\end{equation}
We will perform the quantization of this Hamiltonian \eqref{eq:H_density} in Sec. \ref{sec: CANONICAL QUANTIZATION} and will show that it is Hermitian.

\section{The two-point green's function}
\label{sec: THE TWO-POINT GREEN'S FUNCTION FROM FEYNMAN PATH INTEGRAL}
Two-point Green’s functions play a central role in understanding the causal structure of any field‑theoretic system. In this framework \cite{Galley2013}, which allows both forward and backward evolution, we derive both the advanced and retarded Green’s functions for our scalar dissipative model. We begin with the action $S[\phi_+,\phi_-]$ defined in terms of the Lagrangian density $\Omega(\phi_\pm,\partial_\mu\phi_\pm)$:
\begin{equation}
\begin{aligned}
S[\phi_+,\phi_-] = \int d^4x \Bigl[\partial_\mu \phi_-\,\partial^\mu \phi_+
  - m^2 \,\phi_+ \,\phi_-
  - \gamma\,\phi_-\,\dot{\phi}_+\,\Bigl],
  \label{eg:action in position}
\end{aligned}
\end{equation}
To facilitate the analysis of this dissipative model, the action is expressed in momentum space by performing a Fourier transform of the fields, thereby rendering derivatives as algebraic factors and diagonalizing the quadratic part of the action:
\begin{equation}
    \phi_\pm(x) = \int \frac{d^4k}{(2\pi)^4}\,e^{i k\cdot x}\,\tilde\phi_\pm(k)\,.
\end{equation}
Substituting into the action in Eq.~\eqref{eg:action in position}, one obtains
\begin{equation}
\begin{aligned}
S[\tilde\phi_+,\tilde\phi_-]
&= \int \frac{d^4k}{(2\pi)^4}\,
   \tilde\phi_+(k)\,\bigl[k^2 - m^2 - i\,\gamma\,k^0\bigr]\,
   \tilde\phi_-(-k)\,,
\end{aligned}
\label{eg:scalar dissipative action mom space}
\end{equation}
where $k^2 = (k^0)^2 - \mathbf{k}^2\,$

We rewrite the action in a compact matrix form to make its quadratic structure explicit, allowing the path integral to be expressed as a Gaussian over the field multiplet $\tilde{\Phi}(k)$. This representation isolates all dynamics in the kernel $M(k)$, simplifying the computation of the two-point function and higher-order corrections.

\begin{equation}
S[\tilde{\phi}_+, \tilde{\phi}_-] = \frac{1}{2}\int \frac{d^4k}{(2\pi)^4} \, \tilde\Phi(k)^T M(k) \tilde\Phi(-k)\,,
\label{eg:AA}
\end{equation}
where the matrix $M(k)$ is given by:
\begin{equation}
M(k) = \begin{bmatrix} 
0 & k^2 - m^2 - i \gamma \omega_k \\
k^2 - m^2 + i \gamma \omega_k & 0 
\end{bmatrix}\,,
\end{equation}
and
\begin{equation}
\tilde\Phi^T(k) = \begin{bmatrix} 
\tilde{\phi}_+(k) & \tilde{\phi}_-(k) 
\end{bmatrix}
\quad \text{and} \quad 
\tilde\Phi(-k) = \begin{bmatrix} 
\tilde{\phi}_+(-k) \\
\tilde{\phi}_-(-k) 
\end{bmatrix}\,.
\end{equation}

Since the action in Eq.~\eqref{eg:AA} is explicitly quadratic in the fields with kernel $M(k)$, the path integral reduces to a Gaussian functional integral, which in momentum space can be written as,

\begin{equation}
    \begin{aligned}
    Z[J] & =\int\mathcal{D}\tilde \Phi exp\biggl[\frac{i}{2}\int \frac{d^4k}{(2\pi)^4} \Bigg(
\begin{aligned}
& \tilde\Phi(k)^T M(K)\tilde \Phi(-k) \\
&+ J^T(k) \tilde\Phi(-k) + J^T(-k) \tilde\Phi(+k) 
\end{aligned}
\Bigg)\biggl],\\
&\quad\text{with}\quad
J^T(k)=\bigl[J_+(k),\,J_-(k)\bigr],\\
&= \mathcal{N}\,
\exp\!\biggl[-\frac{i}{2}\int\frac{d^4k}{(2\pi)^4}
J^T(-k)\,M^{-1}(k)\,J(k)\biggr]\,.
\label{eg:path-integral}
\end{aligned}
\end{equation}
The factor $\mathcal{N}$ appearing in Eq.~\eqref{eg:path-integral} is a normalization constant, and $M^{-1}(k)$ is the inverse of the matrix $M(k)$. The inverse matrix thus corresponds to the Green’s function matrix of the theory:
\begin{equation}
\tilde G(k) \equiv M^{-1}(k)
= \begin{pmatrix}
0 & \tilde G_{\mathrm{ret}}(k) \\
\tilde G_{\mathrm{adv}}(k) & 0
\end{pmatrix}\,,
\end{equation}
where the off‑diagonal components are identified as the retarded and advanced Green’s functions:
\begin{equation}
\tilde G_{\mathrm{ret}}(k)
= \frac{1}{k^2 - m^2 + i\,\gamma\,k^0}\,, 
\qquad
\tilde G_{\mathrm{adv}}(k)
= \frac{1}{k^2 - m^2 - i\,\gamma\,k^0}\,.
\label{eg: momentum space greens  function}
\end{equation}
These expressions are given in momentum space. In position space, one obtains
\begin{equation}
G_{\mathrm{ret}}(x-y)
= \int\frac{d^4k}{(2\pi)^4}\,
\frac{1}{k^2 - m^2 + i\,\gamma\,k^0}\,
e^{-i k\cdot (x-y)}\,,\label{eg:two point retarted function}
\end{equation}
and
\begin{equation}
G_{\mathrm{adv}}(x-y)
= \int\frac{d^4k}{(2\pi)^4}\,
\frac{1}{k^2 - m^2 - i\,\gamma\,k^0}\,\label{eg: two point advance function}
e^{-i k\cdot (x-y)}\,.
\end{equation}

\begin{figure}
    \centering
    \begin{subfigure}{0.49\linewidth}
        \centering
        \includegraphics[width=\linewidth]{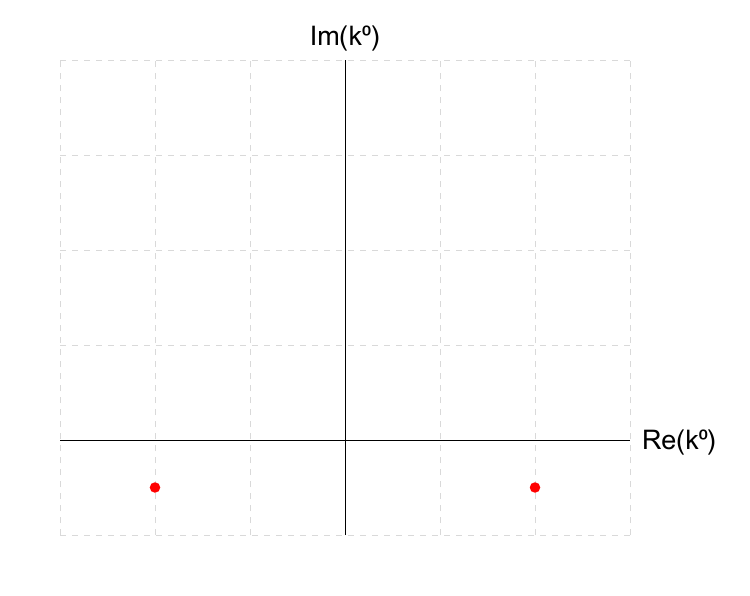}
        \caption{}
    \end{subfigure}%
    \hfill
    \begin{subfigure}{0.49\linewidth}
        \centering
        \includegraphics[width=\linewidth]{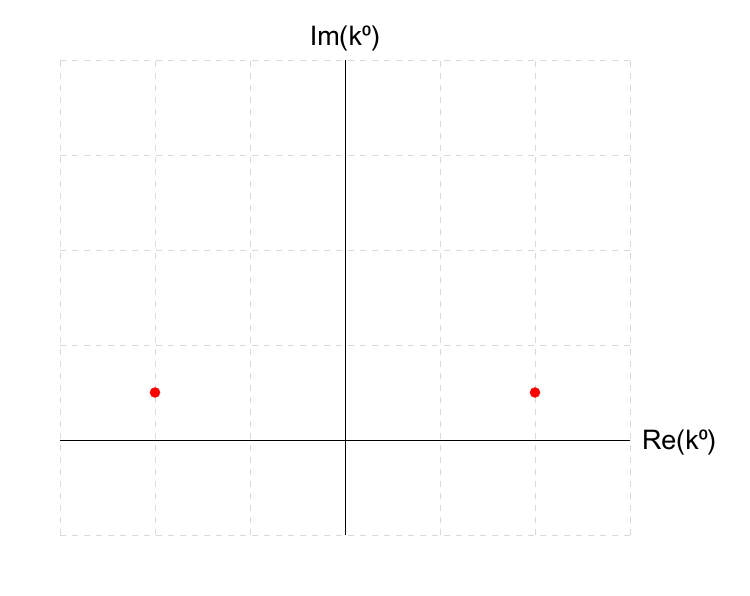}
        \caption{}
    \end{subfigure}
    \caption{(a) Pole of the retarded Green’s function\eqref{eg:two point retarted function}, shifted just below the real axis; (b) pole of the advanced Green’s function\eqref{eg: two point advance function}, shifted just above the real axis.}
    \label{fig: poles of Greens function}
\end{figure}

The above two equations yield the retarded and advanced two-point Green’s functions of the real scalar dissipative field. The poles of the retarded (advanced) Green’s function are shifted slightly below (above) the real $k^0$ axis, as illustrated in Fig.~\ref{fig: poles of Greens function}.  

In contrast to standard QFT, the dissipative term in the action naturally introduces the $i\gamma$  factor into the two-point function. The expressions for the retarded and advanced Green’s functions in Eqs.~\eqref{eg:two point retarted function} and \eqref{eg: two point advance function} are similar to the results reported in Refs.~\cite{Sieberer2016,Baggioli2020}.

\subsection{Spectral Function}
To interpret the poles in Eq.~\eqref{eg: momentum space greens function}, we compute the spectral function for the scalar dissipative theory. Following the definition in Refs.~\cite{Jacob:1961zz, Kuksa:2015iaa, Giacosa:2021mbz}, the spectral function is
\begin{equation}
\rho(k^2)
= -\frac{1}{\pi}\,\Im\!\bigl[\tilde G_{\mathrm{ret}}(k)\bigr]
= \frac{1}{\pi}\,
  \frac{\gamma\,k^0}
       {\bigl[(k^0)^2 - E_{\vec{k}}^2\bigr]^2 + \gamma^2\,(k^0)^2}\,,
\label{eg:spectral_function}
\end{equation}
where $E_{\vec{k}}^2 = \mathbf{k}^2 + m^2$.  

The spectral function peaks near \(k^0 \approx E_{\vec{k}}\) with a width \(\delta \approx \gamma\), corresponding to a particle lifetime \(\tau \sim 1/\gamma\). We refer to this emergent excitation as a quasi‑scalar boson, since it is not fundamental but arises due to the non‑conservative interaction in the theory. The spectral function satisfies \(\rho(k^2)\xrightarrow[]{}\operatorname{sgn}(k^0)\,\delta\bigl(k^2 - m^2\bigr)\) in the limit \(\gamma\xrightarrow{}0\), which exactly matches the spectral function of free scalar field theory. Equivalently, by setting \(\gamma=0\) (i.e., switching off dissipation), the retarded and advanced Green’s functions in Eqs.~\eqref{eg:two point retarted function} and \eqref{eg: two point advance function} coincide with the standard Feynman Green's function of the free theory.

\section{CANONICAL QUANTIZATION}
\label{sec: CANONICAL QUANTIZATION}
In this section, we perform the canonical quantization of the model and construct the Fock-space representation of the Hamiltonian in Eq.~\eqref{eq:H_density}. Starting from the Lagrangian density in Eq.~\eqref{eg:pm L} and substituting into the Euler--Lagrange equation, Eq.~\eqref{eg:eq_of_motion_before_Pl}, we obtain the field equations:
\begin{align}
    \partial_\mu \partial^\mu \phi_+ + m^2 \phi_+ + \gamma\,\dot\phi_+ &= 0\,, 
    \label{eg:eq_of_motion_plus} \\
    \partial_\mu \partial^\mu \phi_- + m^2 \phi_- - \gamma\,\dot\phi_- &= 0\,.
    \label{eg:eq_of_motion_minus}
\end{align}

These field equations differ from the usual Klein-Gordon form due to the presence of damping and anti-damping terms. The first equation, for $\phi_+$, describes the physical damping of the system, whereas the second equation, for $\phi_-$, describes anti-damping that absorbs and encodes the dissipation appearing in the first equation. It should be emphasized that, in the classical physical limit, only the first equation contributes while the second equation vanishes. Nevertheless, the damped(anti-damped) plane-wave ansatz,
\begin{equation}
    \phi_+(x) = \mathcal{A}\,e^{-\frac{\gamma}{2}t}e^{-i k\cdot x}\,, 
    \qquad
    \phi_-(x) = \mathcal{B}\,e^{+\frac{\gamma}{2}t}e^{-i k\cdot x}
    \label{eg:damped_ansatz}
\end{equation}
satisfies Eqs.~\eqref{eg:eq_of_motion_plus} and \eqref{eg:eq_of_motion_minus}. Since each solution splits into an overall exponential factor and a standard plane wave, we can use the plane‑wave part to perform the usual mode expansion. In the Heisenberg picture the fields become,

\begin{align}
\phi_+(x)
&= e^{-\tfrac{\gamma}{2}t}
  \int\!\frac{d^3\mathbf{k}}{(2\pi)^3\sqrt{\omega_k}}
  \Bigl[b(\mathbf{k})\,e^{-i k\cdot x} + b^\dagger(\mathbf{k})\,e^{\,i k\cdot x}\Bigr], 
  \label{eg:mode_plus}\\[6pt]
\phi_-(x)
&= e^{+\tfrac{\gamma}{2}t}
  \int\!\frac{d^3\mathbf{k}}{(2\pi)^3\sqrt{\omega_k}}
  \Bigl[a(\mathbf{k})\,e^{-i k\cdot x} + a^\dagger(\mathbf{k})\,e^{\,i k\cdot x}\Bigr].
  \label{eg:mode_minus}
\end{align}

where $\omega_{\vec{k}}^2=\vec{k}^2+m^2-\frac{\gamma^2}{4}$

From Eq.~\eqref{eg:+-momentum of scalar field} we obtain the conjugate momenta and impose the equal‑time canonical commutation relations, namely
\begin{align}
\bigl[\phi_+(\mathbf{x},t),\,\Pi_-(\mathbf{y},t)\bigr] &= i\,\delta^3(\mathbf{x}-\mathbf{y})\,, 
    \label{eq:phi+_Pi-} \\
\bigl[\phi_-(\mathbf{x},t),\,\Pi_+(\mathbf{y},t)\bigr] &= i\,\delta^3(\mathbf{x}-\mathbf{y})\,.
    \label{eq:phi-_Pi+}
\end{align}
The equal‐time field commutators in Eq.~\eqref{eq:phi+_Pi-} and \eqref{eq:phi-_Pi+} imply that the $\phi_+$ and $\phi_-$ mode operators must mix.  In particular, the only nonzero operator commutators are
\begin{align}
[a(\mathbf{k}),\,b^\dagger(\mathbf{p})] &= (2\pi)^3\,\delta^3(\mathbf{k}-\mathbf{p})\,, 
    \label{eq:abdag} \\
[b(\mathbf{k}),\,a^\dagger(\mathbf{p})] &= (2\pi)^3\,\delta^3(\mathbf{k}-\mathbf{p})\,,
    \label{eq:badag}
\end{align}
and all other commutators vanish:
\begin{equation}
\begin{aligned}
[a(\mathbf{k}),a^\dagger(\mathbf{p})]
&=[b(\mathbf{k}),b^\dagger(\mathbf{p})]
=[a(\mathbf{k}),a(\mathbf{p})]
=[b(\mathbf{k}),b(\mathbf{p})] \\
&=[a^\dagger(\mathbf{k}),a^\dagger(\mathbf{p})]
=[b^\dagger(\mathbf{k}),b^\dagger(\mathbf{p})]
=0\,.
\label{eq:vanishing relation}
\end{aligned}
\end{equation}

The scalar field operator $\phi_+$ in Eq.~\eqref{eg:mode_plus} is identified as the physical field of the theory, defined as an average over the forward- and backward-evolving fields via the transformation in Eq.~\eqref{eg:newphitrans}. The operator $\phi_+$ creates physical excitations, which we previously mentioned as quasi-scalar bosons. These excitations decay as $\exp(-\gamma t/2)$, thereby encoding dissipation in the system. They serve as the quantum analog of the classical physical-limit condition~\cite{Galley2013}. Since $\phi_+$ is an operator-valued field in quantum field theory, it is not meaningful to impose the physical-limit condition directly at the operator level. Instead, we define the physical state corresponding to the operator $\phi_+$. In contrast, the scalar field operator $\phi_-$ in Eq.~\eqref{eg:mode_minus} plays the role of an auxiliary field, whose excitations are not directly observable. Nevertheless, $\phi_-$ appears in physical correlation functions and grows as $\exp(\gamma t/2)$, compensating for the decay of $\phi_+$. 
 
An analogous interpretation holds for the creation and annihilation operators: $b^\dagger(k)$ and $b(k)$  in Eq.~\eqref{eg:mode_plus} create and annihilate physical momentum states, whereas $a^\dagger(k)$ and $a(k)$ in Eq.~\eqref{eg:mode_minus} serve as auxiliary creation and annihilation  operators that create and annihilate Partner excitation which is required to normalized the  physical momentum states.

\subsection{Fock space representation of Hamiltonian}
In this subsection, we find the Fock space representation of the classical Hamiltonian. We start with Eq.~\eqref{eq:H_density}, the classical Hamiltonian for the scalar dissipative model. By substituting Eq.~\eqref{eg:+-momentum of scalar field} into this expression, we obtain
\begin{equation}
    H = \int d^3x\,\bigl(\dot{\phi}_+\dot{\phi}_- + \nabla\phi_+\cdot\nabla\phi_- + m^2\phi_+\phi_-\bigr)
    \label{eq:H1}
\end{equation}
To derive the Fock-space representation of the Hamiltonian~\eqref{eq:H1}, we insert the mode expansions from Eqs.~\eqref{eg:mode_plus} and \eqref{eg:mode_minus} into \eqref{eq:H1}, yielding
\begin{equation}
    H = \int \frac{d^3p}{(2\pi)^3}
    \Bigl[\bigl(\omega_{\vec p} + i\tfrac{\gamma}{2}\bigr)\,b^\dagger(\vec p)\,a(\vec p)
    + \bigl(\omega_{\vec p} - i\tfrac{\gamma}{2}\bigr)\,b(\vec p)\,a^\dagger(\vec p)\Bigr].
\end{equation}

Applying the commutation relation \eqref{eq:badag} leads to
\begin{equation}
    H = \int \frac{d^3p}{(2\pi)^3}
    \Bigl[\bigl(\omega_{\vec p} + i\tfrac{\gamma}{2}\bigr)\,b^\dagger(\vec p)\,a(\vec p)
    + \bigl(\omega_{\vec p} - i\tfrac{\gamma}{2}\bigr)\,a^\dagger(\vec p)\,b(\vec p)
    + (2\pi)^3\delta(0)\,\bigl(\omega_{\vec p} - i\tfrac{\gamma}{2}\bigr)\Bigr],
    \label{eq:H3}
\end{equation}
where the term proportional to $\delta(0)$ represents the infinite zero-point energy. By normal ordering (i.e., discarding this divergent vacuum contribution), the Hamiltonian becomes
\begin{equation}
    :H: = \int \frac{d^3p}{(2\pi)^3}
    \Bigl[\bigl(\omega_{\vec p} + i\tfrac{\gamma}{2}\bigr)\,b^\dagger(\vec p)\,a(\vec p)
    + \bigl(\omega_{\vec p} - i\tfrac{\gamma}{2}\bigr)\,a^\dagger(\vec p)\,b(\vec p)\Bigr].
    \label{eq:H4}
\end{equation}
The double Hamiltonian in Eq.~\eqref{eq:H4} is Hermitian, and the imaginary component $\mp\frac{\gamma}{2}$, describes exponential decay(or growth) associated with dissipation.

\subsection{States and normalization condition}
In this subsection, we introduce the physical and auxiliary states and specify their normalization.  As noted above, the physical states are associated with the operators \(b^\dagger\) and \(b\), while the auxiliary states are associated with \(a^\dagger\) and \(a\).  Although the auxiliary states do not correspond to directly observable excitations, they enter the normalization alongside the physical states and serve purely as bookkeeping devices .We begin by defining the normalized vacuum via
\begin{equation}
    b(\vec p)\ket{0} = a(\vec p)\ket{0} = 0.
    \label{eq:vacuum}
\end{equation}
The one‐particle physical states are then generated by acting on the vacuum with \(b^\dagger\):
\begin{equation}
    \ket{p}_{+} = b^\dagger(\vec p)\ket{0}.
    \label{eq:1particle-physical}
\end{equation}
Similarly, the one‐particle auxiliary states are created by
\begin{equation}
    \ket{k}_{-} = a^\dagger(\vec k)\ket{0}.
    \label{eq:1particle-auxiliary}
\end{equation}
These one‐particle states, Eqs.~\eqref{eq:1particle-physical} and \eqref{eq:1particle-auxiliary}, obey the normalization condition
\begin{equation}
   \prescript{}{-}{\langle \vec k \mid \vec p \rangle}_{+}
   = (2\pi)^3 \,\delta^3(\vec k - \vec p)\,.
   \label{eq:normalization}
\end{equation}
Equation~\eqref{eq:normalization} shows that the auxiliary states contribute on an equal footing to the overall normalization.  By repeatedly applying the physical‐state creation operators \(b^\dagger\) to the vacuum, one constructs an \(n\)‐particle physical state:
\begin{equation}
    \ket{\vec p_1, \ldots, \vec p_n}_{+}
    = b^\dagger(\vec p_1)\,b^\dagger(\vec p_2)\,\cdots\,b^\dagger(\vec p_n)\,\ket{0}\,.
    \label{eq:nparticle-physical}
\end{equation}
In Eq.~\eqref{eq:nparticle-physical}, all the \(b^\dagger\) operators commute (see Eq.~\eqref{eq:vanishing relation}), so swapping any two momenta doesn’t change the state.  This tells us that these particles are bosons.

To show that these bosons decay over time, we use the Heisenberg equation of motion\cite{Peskin:1995ev}
\begin{equation}
    \frac{d\,b^\dagger(\vec k,t)}{dt}
    = i\bigl[H,\,b^\dagger(\vec k,t)\bigr].
\end{equation}
Plugging in our Hamiltonian from Eq.~\eqref{eq:H4}, we find
\begin{equation}
    \frac{d\,b^\dagger}{dt}
    = i\!\bigl(\omega_{\vec k} + i\,\tfrac{\gamma}{2}\bigr)\,b^\dagger.
\end{equation}
Solving this gives
\begin{equation}
    b^\dagger(\vec k,t)
    = e^{\,i\omega_{\vec k} t}\;e^{-\frac{\gamma}{2}t}\;b^\dagger(\vec k,0).
    \label{eq: b in hesisengber picture}
\end{equation}
Using Eq.~\eqref{eq: b in hesisengber picture}, the time evolution of the one‐particle physical state is
\begin{align}
    \ket{\vec{k},t}_{+}
    &= b^\dagger(\vec{k},t)\,\ket{0}
    \nonumber\\
    &= e^{\,i\omega_{\vec{k}}t}\;e^{-\frac{\gamma}{2}t}\;\ket{\vec{k},0}_{+}\,.
    \label{eq:time-evolution-+}
\end{align}
Hence, from Eq.~\eqref{eq:time-evolution-+}, the physical state decays exponentially in time, with an average lifetime $\tau \sim \frac{1}{\gamma}$.

Similarly, the one‐particle auxiliary state evolves as
\begin{equation}
    \ket{\vec{k},t}_{-}
    = a^\dagger(\vec{k},t)\,\ket{0}
    = e^{\,i\omega_{\vec{k}}t}\;e^{\frac{\gamma}{2}t}\;\ket{\vec{k},0}_{-}\,.
    \label{eq:time-evolution--}
\end{equation}
The overlap of these time‐evolved one‐particle states is
\begin{equation}
    \prescript{}{-}{\bigl\langle \vec{k},t \,\bigm|\, \vec{p},t \bigr\rangle}_{+}
    = (2\pi)^3 \,e^{-\,i\bigl(\omega_{\vec p}-\omega_{\vec k}\bigr)t}\,
      \delta^3(\vec k - \vec p)\,.
    \label{eq:normalized-time-evolution}
\end{equation}
This shows that probability is conserved, as expected for a Hermitian Hamiltonian in Eq.~\eqref{eq:H4}. 

\section{Conclusions}
We performed the first-principles quantization of a non-conservative scalar field model with linear damping using the action principle proposed by Galley~\cite{Galley2013}. A non-conservative potential term is \emph{constructed} in  Eq.~\eqref{eg:dissipativelagrangian} to describe linear dissipation, which leads to a damped Klein-Gordon equation within this framework. Using the path integral formalism, we derived the two-point Green's function for the model and demonstrated that dissipative interactions shift the poles of the propagator away from the real axis. The pole shifted below the real axis corresponds to the retarded Green's function, while the pole shifted above corresponds to the advanced Green's function. We also perform canonical quantization of the theory and find that the resulting dissipative scalar field system is governed by a "double hermitian" Hamiltonian. Our results are consistent with the standard results in the Schwinger-Keldysh in-in formalism based on the \emph{effective field theoretic} formulation of quantum field theory~\cite{Sieberer2016, Baggioli2020}. 

We distinguish between physical and auxiliary states: the physical state arises from the field in Eq.~\eqref{eg:mode_plus}, which creates quasi-scalar excitations in the vacuum, whereas the auxiliary state originates from the field in Eq.~\eqref{eg:mode_minus} and produces the counterpart of the quasi-scalar excitation. The auxiliary state contributes only to the normalization of the states and has no physical significance. We show that the quasi-scalar boson decays over time and possesses a finite lifetime.

Ab initio constructions of non-conservative field theoretic models not only draws connections between classical and sophisticated quantum field theoretic formalisms, but also presents an opportunity to investigate potential classical and semi-classical implications in the context of cosmology and effective quantum gravity. Schwinger-Keldysh (or in-in) formalism has been applied to inflationary dynamics in the past to resolve fundamental inconsistencies with QFT on a dynamical gravitational background. As future projects, relevant models with dissipative scalar and tensor fields will be developed to study classical and semi-classical effects of such fields in the early universe cosmology.

\bibliographystyle{apsrev4-2}
\bibliography{references}

\end{document}